%% file: paper8.tex
\documentclass[12pt]{iopart}
\usepackage[dvips]{graphics}
\usepackage{epic,jump}
\newcommand{\ZZ}{{\mathbf{Z}}}
\newcommand{\NN}{{\mathbf{N}}}
\begin{document}
\jl{1}
\title[Density conserving CA]{Cellular  automaton rules conserving the
number of active sites}
\author{Nino Boccara\dag { }and Henryk Fuk\'s\ddag}
\address{University of Illinois at Chicago, Dept. of Physics, Chicago,
IL 60607-7059, USA}
\address{\dag\tt boccara@uic.edu}
\address{\ddag\tt fuks@sunphy1.phy.uic.edu}

\begin{abstract}
This paper shows how to determine all the unidimensional two-state
cellular automaton rules of a given number of inputs which conserve
the number of active sites. These rules have to satisfy a necessary
and sufficient condition. If the active sites are viewed as cells
occupied by identical particles, these cellular automaton rules
represent evolution operators of systems of identical interacting
particles whose total number is conserved. Some of these rules, which
allow motion in both directions, mimic ensembles of one-dimensional
pseudo-random walkers. Numerical evidence indicates that the
corresponding stochastic processes might be non-Gaussian.
\end{abstract}
\submitted
\maketitle

\section{Introduction}
Systems which consist of a large number of simple identical elements
evolving in time according to simple rules often exhibit a complex
behavior as a result of the cooperative effect of their components.
Cellular automata are models of such systems. They may be defined as
follows: Let $s:\ZZ\times\NN\mapsto\{0,1\}$ be a function that
satisfies the equation
\begin{equation}
s(i,t+1)=f\big(s(i-r_l,t),s(i-r_l+1,t),\ldots,s(i+r_r,t)\big),
\label{g-rule}
\end{equation}
for all $i\in\ZZ$ and all $t\in\NN$, where $\ZZ$ is the set of all
integers and $\NN$ the set of nonnegative integers.  Such a discrete
dynamical system is a two-state one-dimensional cellular automaton
(CA). The mapping $f : \{0,1\}^{r_l+r_r+1}\to\{0,1\}$ is the rule, and
the positive integers $r_l$ and $r_r$ are, respectively, the left and
right radius of the rule. $f$ will also be called an $n$-input rule
where $n=r_l+r_r+1$. The function $S_t:i\mapsto s(i,t)$ is the state
of the CA at time $t$. ${\cal S}=\{0,1\}^{\ZZ}$ is the state space. An
element of the state space is also called a configuration. Since the
state $S_{t+1}$ at time $t+1$ is entirely determined by the state
$S_t$ at time $t$ and the rule $f$, there exists a unique mapping
$F_f:{\cal S}\to{\cal S}$ such that $S_{t+1}=F_f(S_t)$. $F_f$, which
is the evolution operator, is also referred to as the global CA rule.

CA have been widely used to model complex systems in which the local
character of the rule plays an essential r\^ole (Wolfram 1983, Farmer
{\it et al\/} 1984, Manneville {\it et al\/} 1989, Gutowitz 1990,
Boccara {\it et al\/} 1993). In the past few years, CAs have been
successfully used to model highway traffic. One of the simplest model
is defined on a one-dimensional lattice of $L$ sites with periodic
boundary conditions. Each site is either occupied by a vehicle, or
empty. The velocity  of each vehicle is an integer between 0 and
$v_{\rm max}$. If $x(i,t)$ denotes the position of car $i$ at time
$t$, the position of the next car ahead at the same time is
$x(i+1,t)$. With this notation, the system evolves according to a
synchronous rule given by
\begin{equation}
x(i,t+1)=x(i,t)+v(i,t+1),
\end{equation}
where
\begin{equation}
v(i,t+1)=\min\big(x(i+1,t)-x(i,t)-1,x(i,t)-x(i,t-1)+a,v_{\rm max}\big)
\end{equation}
is the velocity of car $i$ at time $t+1$. $x(i+1,t)-x(i,t)-1$ is the
gap (number of empty sites) between cars $i$ and $i+1$ at time $t$,
$x(i,t)-x(i,t-1)$ is the velocity $v(i,t)$ of car $i$ at time $t$, and
$a$ is the acceleration. $a=1$ corresponds to the deterministic model
of Nagel and Schreckenberg (1992) while the case $a=v_{\rm max}$ has
been considered by Fukui and Ishibashi (1995). In this last case, the
evolution rule can be written
\begin{equation}
x(i,t+1)=x(i,t)+\min\big(x(i+1,t)-x(i,t)-1,v_{\rm max}\big).
\label{a_max_rule}
\end{equation}
This is a cellular automaton rule with, at least, its left radius
equal to $v_{\rm max}$ and its right one equal to $v_{\rm max}-1$. The
case $a<v_{\rm max}$ is a second order rule, that is, the state at
time $t+1$ depends upon the states at times $t$ and $t-1$. For $v_{\rm
max}=1$,  these two rules coincide with elementary CA rule 184 (rule
code numbers as in Wolfram 1994)

Since, for these highway traffic models on a ring (we shall always
consider cyclic boundary conditions), the number of cars is conserved,
it might be interesting to address the more general question: Is it
possible to determine all one-dimensional two-state CA rules which
conserve the number of active sites? We cannot expect that all these
rules will mimic realistic highway traffic. It is preferable to
view them as describing the evolution of systems which consist of a
fixed number of interacting particles.

\section{General considerations}
If the sites are either all inactive or all active, they should remain
so during the evolution. Therefore, for any number of inputs $n$, the
local rule should satisfy the conditions
\begin{eqnarray}
f(\underbrace{0,0,0,\ldots,0}_n) &=& 0 \label{cond-0} \\
f(\underbrace{1,1,1,\ldots,1}_n) &=& 1 \label{cond-1}.
\end{eqnarray}

If the rule~(\ref{g-rule}) changes the site value $s(i,t)$, we may say
that it either ``created'' a particle, if $s(i,t+1)=0$ and
$s(i,t+1)=1$, or ``annihilated'' a particle in the opposite case.
Since, the argument $s(i,t)$ of function $f$ takes the values 0 and 1
an equal number of times, conservation of particles number implies
that the number of creations and annihilations should be equal. In
other words, the number of preimages of 0 and 1 by $f$ should be the
same.

Consider rules $f_1$ and $f_2$, whose radii are, respectively,
$r_{l1}$, and $r_{r1}$, and $r_{l2}$ and $r_{r2}$. The rule $f_1\circ
f_2$ which consists, at each time step, in the successive application
of $f_1$ and $f_2$, conserves the number of particles if $f_1$ and
$f_2$ do. Its radii are $r_l=r_{l1}+r_{l2}$ and $r_r=r_{r1}+r_{r2}$.
For instance, the 4-input rule whose binary code number is
1011100010111000 ($r_l=1$, $r_l=2$) conserves the number of particles
since it is the composition of the left shift (binary code number
1010, $r_l=0$, $r_r=1$ ) and rule 184 (binary code number 10111000,
$r_l=r_r=1$) which both conserve the number of particles.

If, as for highway traffic, we wish to follow particles motion, it
might be useful to define a representation of rule $f$ which exhibits
this motion. Such a ``motion representation'' may be defined as
follows. List all the neighbourhoods of a given particle represented
by $\bf 1$. Then, for each neighbourhood, indicate the displacement of
this particle by an integer $v$, where $v$ is positive if the particle
moves to the right and negative if it moves to the left. For instance,
the motion representation of Rule 184 would be
\begin{equation}
{{\bf 1}0}\ \ 1,\quad {{\bf 1}1}\ \ 0.
\end{equation}
Since, for this particular rule, the particle can only move to the
right, we only need to indicate the relevant neighbourhood of the
particle. This representation can be made more visual if we draw an
arrow joining the initial and final positions of the particle, i.e.,
for Rule 184
\begin{equation}
\mbox{\jumponeright{10}{2}{0}}, \quad
\mbox{\zerojump{11}{2}{0}}.
\end{equation}
Note that, in this case, it is not necessary to specify the moving
particle by a bold digit.

This last notation is very compact. For instance, the 4-input rule
which results from the composition of Rule 184 and the left shift, is
represented by
\[
\mbox{\zerojump{10}{2}{0}}, \quad
\mbox{\jumponeleft{$\bullet$11}{3}{1}}.
\]
where $\bullet$ represents either 0 or 1. The motion representation
has another advantage. When we are interested by the motion of the
particles, the knowledge of the rule table, which gives the images of
the various $n$-inputs, is not sufficient. We have to specify the
values of the right and left radii since modifying $r_l$ and $r_r$ at
constant $n$ is equivalent to adding a constant velocity to all the
particles.

Rules obtained by reflection or conjugation of a rule conserving the
number of active sites have the same property. Reflection exchanges
the values of  $r_l$ and $r_r$ and changes the sign of the velocity.
Conjugation exchanges the r\^oles of 0s and 1s, that is, if a rule
describes a specific motion of particles (represented by 1s) then its
conjugate describes the same rule,  but for the motion of holes
(represented by 0s). If $R$ and $C$ denote, respectively, these two
operators, two $n$-input rules $f_1$ and $f_2$ are said to be
equivalent if there exists an element $g$ of the four-group generated
by $R$ and $C$ which transforms $f_1$ into $f_2$.

\section{Rules determination}
One method to determine all the $n$-input rules $f$ conserving the
number of active sites is to find a system of equations whose
solutions are all the functions
\begin{equation}
f:\{0,1\}^n\mapsto\{0,1\} \label{n-sol}
\end{equation}
which, for all $L\ge n$, satisfy the conditions
\begin{eqnarray}
f(x_1,x_2,\ldots,x_n) + f(x_2,x_3,\ldots,x_{n+1}) + \cdots  +
f(x_L,x_1,\ldots,x_{n-1})  \nonumber \\
\qquad= x_1+x_2+\cdots+x_L,
\label{L-cond}
\end{eqnarray}
for all $L$-ring configurations (cyclic permutations). Such a system
will be called an $L$-system of equations. Conditions~(\ref{L-cond}) are
clearly necessary, but does it exist a minimum value $L_{\rm min}$ of
$L$ such that they are also sufficient?

We shall prove that  $L_{\rm min}$ {\sl exists, and is equal to\/}
$2n-2$. That is, {\it the necessary and sufficient condition for a
rule $f$ to conserve the number of active sites is to satisfy
Relations~(\ref{L-cond}) for $L=2n-2$.}

Before giving a formal proof of this result, we shall present a
simple, but not rigorous, argument.
Given the states $s(1,t),s(2,t),\ldots,s(n,t)$ of sites $1,2,\ldots,n$
at time $t$, the state $s(r_{r_l+1},t+1)$ of site $r_l+1$ at time
$t+1$ is determined ($n=r_l+r_r+1$). To determine the states of sites
1 and $n$ at time $t+1$, we also need to know the states at time $t$
of the $r_l$ sites on the left of site 1 and the $r_r$ sites on the
right of site $n$. To obtain the minimum number of sufficient
conditions satisfied by~(\ref{n-sol}), we shall require that the minimum
number of sites we have to add to the original $n$ sites should be
such that their state values at time $t+1$ should depend on, at least,
one of the site values $s(1,t),s(2,t),\ldots,s(n,t)$. This condition
implies that we should consider an $L_{\rm min}$-ring in which the
sites $1-r_l$ and $n+r_r$ coincide. Therefore, $L_{\rm
min}=r_l+n+r_r-1$, that is, $L_{\rm min}=2n-2$.

To prove the above result in a more rigorous way, we shall show that,
if $L>2n-2$, any equation of an $L$-system is a linear combination of
three equations belonging, respectively, to $(L-1)$-, $(2n-3)$-, and
$(2n-2)$-systems. More precisely, for all $L$-ring configurations
$\{x_1,x_2,\ldots,x_L\}$, Equation~(\ref{L-cond}) can be written
\begin{eqnarray}
 & &\Big(f(x_1,x_2,\ldots,x_n) + f(x_2,x_3,\ldots,x_{n+1}) + \cdots
 + f(x_{L-1},x_1,\ldots,x_{n-1})\Big) \nonumber \\
 & & -\Big(f(x_1,x_2,\ldots,x_{n-2},x_{L-n+1},x_{L-n+2})\nonumber \\
 & & + f(x_2,x_3,\ldots,x_{L-n+3}) + \cdots +
 f(x_{n-2},x_{L-n+1},\ldots,x_{L-1})\nonumber \\
 & & + f(x_{L-n+1},x_{L-n+2},\ldots,x_{L-1},x_1) + \ldots +
 f(x_{L-1},x_1,\ldots,x_{n-2},x_{L-n+1})\Big)\nonumber \\
 & & +\Big(f(x_1,x_2,\ldots,x_{n-2},x_{L-n+1},x_{L-n+2})+
 f(x_2,x_3,\ldots,x_{L-n+3}) + \cdots \nonumber \\
 & &+
 f(x_{n-2},x_{L-n+1},\ldots,x_{L-1}) +
 f(x_{L-n+1},x_{L-n+2},\ldots,x_{L-1},x_L) + \ldots\nonumber \\
 & &+
 f(x_{L-1},x_L,x_1,\ldots,x_{n-2}) +
 f(x_L,x_1,\ldots,x_{n-2},x_{L-n+1})\Big)\nonumber \\
 & & =
 (x_1+x_2+\cdots+x_{L-1})
 - (x_1+\cdots+x_{n-2}+x_{L-n+1}+\cdots+x_{L-1})\nonumber \\
 & &+ (x_1+\cdots+x_{n-2}+x_{L-n+1}+\cdots+x_L).
\label{eqn-decomp}
\end{eqnarray}

To verify this result, we have to assume that $x_{L-1}=x_L$, which is
always the case for any cycle, except, when $L$ is even, for the cycle
$1010\ldots 10$. Verifying (\ref{eqn-decomp}) is then a bit tedious but
straightforward. By induction, relation (\ref{eqn-decomp}) shows that
any equation of an $L$-system is a linear combination of equations
belonging to $(2n-3)$-, and $(2n-2)$-systems.

The equation corresponding to the cyclic configuration $1010\ldots 10$
reads
\begin{equation}
\underbrace{f(1010\ldots 10)+f(0101\ldots 01)+\cdots+f(0101\ldots 01)}_L
= \frac{L}{2}
\end{equation}
if $n$ is even, and
\begin{equation}
\underbrace{f(1010\ldots 01)+f(0101\ldots 10)+\cdots+f(0101\ldots 10)}_L
= \frac{L}{2}
\end{equation}
if $n$ is odd.
That is,
\begin{equation}
f(1010\ldots 10)+f(0101\ldots 01) = 1
\end{equation}
if $n$ is even, and
\begin{equation}
f(1010\ldots 01)+f(0101\ldots 10) = 1
\end{equation}
if $n$ is odd. One of the images by $f$ of the two alternating
$n$-sequences of 0s and 1s is equal to 1, and the other one to 0.

\section{Examples}
One- and two-input  rules conserving the number of active sites are
trivial. The identity, represented by
\zerojump{1}{1}{0},
is the only one-input rule, and the left and right shifts, represented
respectively by
\jumponeleft{$\bullet$1}{2}{1} and
\jumponeright{1$\bullet$}{2}{0}, are the only two-input rules.
Note that the rule represented by
\zerojump{1$\bullet$}{2}{0} or
\zerojump{$\bullet$1}{2}{1} is the
identity viewed as a two-input rule, but in agreement with our
convention to only represent the relevant neighbourhood, we shall
always represent it as a one-input rule. This is a general feature.
When we solve the system of equations~(\ref{L-cond}) for $n=3$ and
$L_{\rm min}=4$, we will re-obtain the identity, and the left- and
right-shifts as 3-input rules.

\subsection{3-input rules}
The only nontrivial 3-input rules conserving the number of active sites are
Rules 184 and 226, represented respectively by
\begin{equation}
\mbox{\jumponeright{10}{2}{0}},
\mbox{\zerojump{11}{2}{0}}, \quad\mbox{and}\quad
\mbox{\jumponeleft{01}{2}{1}},
\mbox{\zerojump{11}{2}{1}}.
\end{equation}
Rule 226, which can be obtained either by reflection or conjugation of
Rule 184, models exactly the same deterministic highway traffic rule.
The only difference, clearly shown by the motion representation, is that
cars move to the right instead of moving to the left.

\subsection{4-input rules}
The system of equations~(\ref{L-cond}) for $n=4$ and $L_{\rm min}=6$ has 22
solutions. Among these, we re-obtain the identity, the left- and right-shifts,
Rules 184 and 226 and some simple combinations of these rules viewed as
4-input rules. The new rules are:
\begin{itemize}
\item Rules 43944, 65026, 59946, 49024.  The motion representation
of Rule 43944 ($r_l=2, r_r=1$) is
\[
 \mbox{\jumptworight{100}{3}{0}},
 \mbox{\jumponeright{101}{3}{0}},
 \mbox{\zerojump{11}{2}{0}}.
\]
This rule coincide with the highway traffic rule~(\ref{a_max_rule}) for
$v_{\rm max}=2$, and cars moving to the right. Rule 65026, obtained by
reflection of 43944, describes the same highway traffic rule but for cars
moving in the opposite direction.

The motion representation of Rule 59946, which is the conjugate of
Rule 43944, is
\[
 \mbox{\jumponeleft{011}{3}{2}},
 \mbox{\jumponeleft{101}{3}{2}},
 \mbox{\zerojump{111}{3}{2}}.
\]
It describes a highway traffic rule in which drivers, anticipating
the motion of the car ahead, may move to an occupied site with
$v_{\rm max}=1$. More general rules of this type have been studied
by Fuk\'s and Boccara (1997). Rule 49024 is obtained by reflection
of Rule 59946.

\item Rules 58336, 52930, 63544, 48268.  The motion representation
of Rule 58336 ($r_l=1, r_r=2$) is
\[
 \mbox{\jumponeright{100}{3}{0}},
 \mbox{\zerojump{101}{3}{0}},
 \mbox{\zerojump{11}{2}{0}}.
\]
It describes a highway traffic rule of overcautious drivers who move
to the right with a velocity equal to 1 if, and only if, they have
two empty sites ahead of them.  By reflection we obtain Rule 52930
describing the same highway traffic rule but for cars moving in the
opposite direction.

The motion representation of Rule 63544, conjugate of Rule 58336, is
\[
 \mbox{\jumponeleft{011}{3}{1}},
 \mbox{\zerojump{010}{3}{1}},
 \mbox{\zerojump{11}{2}{1}}.
\]
A particle moves to the left if, and only if, the neighbouring left
site is empty, and the neighbouring right site is occupied. If the
neighbouring left site is occupied the particle does not move. As a
highway traffic rule, it describes drivers who do not like to be
followed, and move to an empty site only when there is a car just
behind them. By reflection we obtain Rule 48268.

\item Rules 56528, 57580, 62660, 51448. The motion representation of Rule
56528 ($r_l=1, r_r=2$) is
\[
 \mbox{\zerojump{100}{3}{0}},
 \mbox{\jumponeright{101}{3}{0}},
 \mbox{\zerojump{11}{2}{0}}.
\]
A particle moves to the right if, and only if, its first right site is
empty and its second right site is occupied. As a highway traffic rule
it describes drivers who move to an empty site if, as a result, they
can be just behind another car. Rule 57580 is obtained by reflection.

The motion representation of Rule 62660, conjugate of Rule 56528, is
\[
 \mbox{\jumponeleft{010}{3}{1}},
 \mbox{\zerojump{011}{3}{1}}.
\]
The particle moves to an empty site on its left if, and only if, there
is an empty site on its right. Rule 51448 is obtained by reflection.

\item{Rules 60200, 48770}. These rules are self-conjugate. The motion
representation of Rule 60200 ($r_l=1, r_r=2$) is
\[
 \mbox{\jumponeright{100}{3}{0}},
 \mbox{\zerojump{101}{3}{0}},
 \mbox{\jumponeleft{011}{3}{1}},
 \mbox{\zerojump{111}{3}{1}}.
\]
A particle moves to the right if, and only if, it has two neighbouring
empty sites on that side. If only the first neighbouring site is empty,
it does not move to avoid occupying a site close to another particle.
If its first right neighbouring site is occupied, then the particle
moves to the left when that site is empty. The effective interaction
between these particles is  repulsive.
Rule 48770, obtained by reflection, describes a similar evolution rule.
\end{itemize}

These last two rules have interesting properties. Starting from a random
initial configuration, after a maximum number of time steps equal to
 $N/2$, where $N$ is the number of sites, the system evolves on its
limit set. This limit
set has a rather simple structure. If the density of particles
$\rho=\frac{1}{2}$, it consists of 3 types of periodic sequences,
namely:
\begin{eqnarray*}
 & &\ldots101010101010\ldots,\quad\hbox{of period 2}\\
 & &\ldots100100100100\ldots,\quad\hbox{of period 3}\\
 & &\ldots110110110110\ldots,\quad\hbox{of period 3}.
\end{eqnarray*}
The probabilities of the various 3-blocks have been determined
numerically. We have found
\begin{eqnarray*}
 & &P(000)=P(111)=0,\\
 & &P(001)=P(110)=P(100)=P(011)=0.145 \pm0,001\\
 & &P(010)=P(101)=0.210\pm0.001.
\end{eqnarray*}

Regarded as a  formal language (Wolfram 1984, Denning  {\it et al\/}
1978, Hopfcroft and Ullmam 1986), such a limit set is regular. Words
in a regular language, on the alphabet $\{0,1\}$, are generated by
walks through a finite directed graph whose arcs are labeled with 0 or
1. Given a finite graph, it is always possible to find an equivalent
deterministic finite graph, that is, a graph in which no more than one
arc of a given label leaves each vertex. For Rules 60200 and 48770,
the corresponding deterministic graph is represented in Figure~1.
\begin{figure}[hb!]
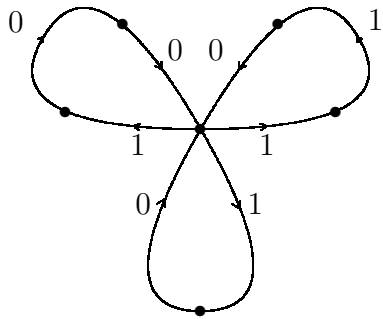

\include{fig1}
\caption{Regular language graph for Rules $60200$ and $48770$.}
\settowidth{\unitlength}{\tt{0}} 
\end{figure}
For $\rho=\frac{1}{2}$, when the CA evolves on its limit set, each
particle performs a pseudo-random walk. The CAs rules being
deterministic, the randomness comes from the randomness of the initial
configuration. Numerical simulations shows that any particle has a
probability $p=0.29$ to move either to the left or to the right, and a
probability $q=1-2p=0.42$ not to move. Actually this pseudo-random
motion is periodic in time, the period being equal to $N/2$.
In the limit set, for a given random initial
configuration, all particles performs the same pseudo-random walk,
with a phase difference depending on the distance separating them.
More precisely, if $X_n(t)$ denotes the position of particle $n$ at
time $t$, for Rule 60200, we have
\[
X_n(t) =X_{n+1}(t-1)-2,
\]
which implies
\[
X_n(t) =X_{n+t}(0)-2t.
\]
This last result shows that the position of a specific particle at
time $t$ is determined by the position of another specific particle
in the initial configuration.

To characterize the nature of the randomness of the
motion of a particle, we have determined the Hurst exponent
(Hurst 1951, Hurst {\it et al\/} 1965, Feder 1988) of the time
series generated
by the displacement of a given particle. Given a time series $s(t)$,
the Hurst exponent $H$ characterizes the asymptotic behaviour of the
standard deviation of $s(t)$ as a function of time. A Brownian motion
(symmetric random walk) has a Hurst exponent $H=\frac{1}{2}$. For a
particle moving according to the 4-input rules 60200 and 48770, we
have found $H=0.63\pm0.02$.
\begin{figure}
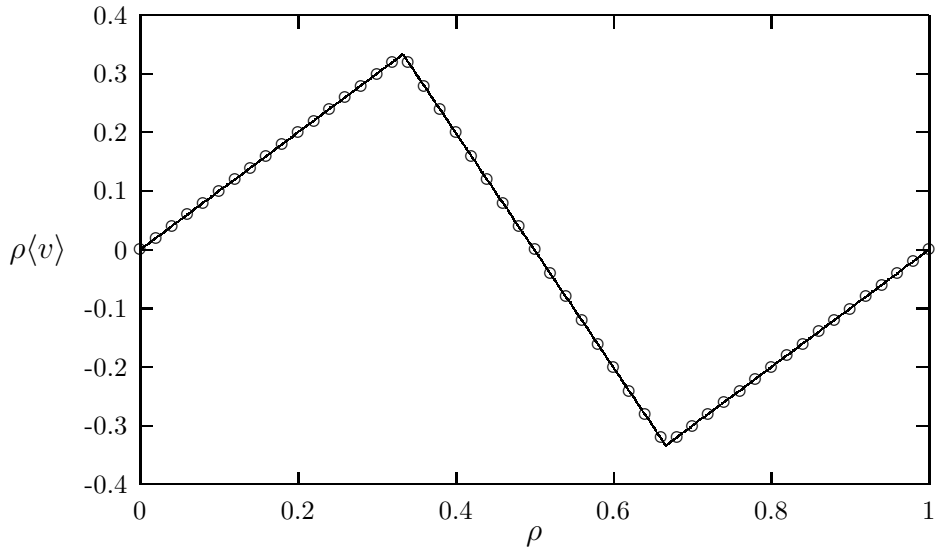

\begin{flushright}
\include{fig2}
\end{flushright}
\caption{Fundamental diagram for Rule $60200$. Small circles represent
numerical results. The piecewise linear line has been obtained using
local structure approximation (see below).}
\settowidth{\unitlength}{\tt{0}} 
\end{figure}
%
Since the pseudo-random motion is periodic in time with a period equal
to half the lattice size, this numerical result is arguable. It could
be interesting to perform a detailed study of the correlations, but,
even correlated random walks may have a Gaussian behavior when the
number of time steps goes to infinity (Weiss 1994). However, in this
case, there exists a crossover between a non-Gaussian and a Gaussian
behavior. This fact implies that, for a large value of the number of
time steps $t$, the exponent of the standard deviation of the walk
could, numerically, be different from 1/2.

When $\rho\ne\frac{1}{2}$, the limit set consists of the previous
periodic  sequences and either sequences of 0s if $\rho<\frac{1}{2}$
or sequences of 1s if $\rho>\frac{1}{2}$. From the motion
representation of Rule 60200, it follows that the average velocity
$\langle v\rangle$ of the particles is $P(100)-P(011)$. The conjugacy
operator changes $\langle v\rangle$ in $-\langle v\rangle$ and $\rho$
in $1-\rho$. Therefore, the so-called ``fundamental diagram'' of road
traffic theory, that is, the graph of the flow $\rho\langle v\rangle$
as a function of the density $\rho$, has a center of symmetry, namely,
the point
$(\rho,\rho\langle v\rangle)=(\frac{1}{2},0)$ (Figure~2). Rule 48770
has identical properties.

\subsection{5-input rules}
The number of rules conserving the number of active sites grows very
fast with the number of inputs. There exist 428 5-input rules
conserving the number of active sites. Few of them are not new either
because they actually depend upon a smaller number of inputs or because
they are simple composition of rules already obtained. In this section
we shall just describe the self-conjugate rules.\footnote{Codes of all
other rules can be obtained from the authors through e-mail.}

There exists 20 self-conjugate rules. Some, such as the identity and
the shifts (left and right, simple and double), are trivial. We also
re-obtain the two self-conjugate 4-input rules. Each of them twice
depending on which side, left or right, the extra input is added.
Finally, we are left with 11 new self-conjugate rules. For
each rule we shall always choose the values of $r_l$ and $r_r$ such that
the condition $\langle v\rangle=0$ for $\rho=\frac{1}{2}$ is satisfied.
This can always be done.

Few of these rules are still not very interesting. After few time steps,
for all $\rho\in]0,1[$, 5 rules emulate the identity, which means that
no particles are moving. These rules are:
Rule 3464560268 ($r_l=1$, $r_r=3$), whose motion
representation is
\[
\mbox{\zerojump{0001}{4}{3}},
\mbox{\zerojump{0111}{4}{2}},
\mbox{\jumponeleft{1001}{4}{3}},
\mbox{\zerojump{0101}{4}{3}},
\mbox{\jumponeright{0110}{4}{2}},
\mbox{\zerojump{111}{3}{2}},
\mbox{\zerojump{1101}{4}{3}},
\]
Rule 3771264248 ($r_l=1$, $r_r=3$), whose motion representation is
\[
\mbox{\zerojump{0011}{4}{2}},
\mbox{\jumponeright{0010}{4}{2}},
\mbox{\zerojump{0011}{4}{3}},
\mbox{\zerojump{0101}{4}{3}},
\mbox{\jumponeleft{1101}{4}{3}},
\mbox{\zerojump{111}{3}{2}},
\mbox{\zerojump{1011}{4}{3}},
\]
and Rule 3824738360 ($r_l=r_r=2$), whose motion representation is
\[
\mbox{\jumponeleft{0011}{4}{2}},
\mbox{\zerojump{0010}{4}{2}},
\mbox{\zerojump{1101}{4}{1}},
\mbox{\zerojump{0111}{4}{2}},
\mbox{\zerojump{101}{3}{2}},
\mbox{\jumponeright{1100}{4}{1}},
\mbox{\zerojump{1111}{4}{2}}.
\]
Rules 4249668928 and 415766320 obtained by reflection of the first
two rules have the same property. Rule 3824738360 is invariant under
reflection.

Rule 3167653058 ($r_l=3$, $r_r=1$), whose motion representation is
\[
\mbox{\jumponeleft{0001}{4}{3}},
\mbox{\zerojump{011}{3}{2}},
\mbox{\zerojump{101}{3}{2}},
\mbox{\zerojump{1111}{4}{2}},
\mbox{\zerojump{1001}{4}{3}},
\mbox{\jumponeright{1110}{4}{2}},
\]
is rather peculiar. As shown in Figure~3, this rule emulates the identity
only for $\rho\in[\frac{1}{3},\frac{2}{3}]$. Rule 4270014080, obtained
by reflection has identical properties. Note that the flow diagram
(Figure~3) is piecewise linear.

\begin{figure}
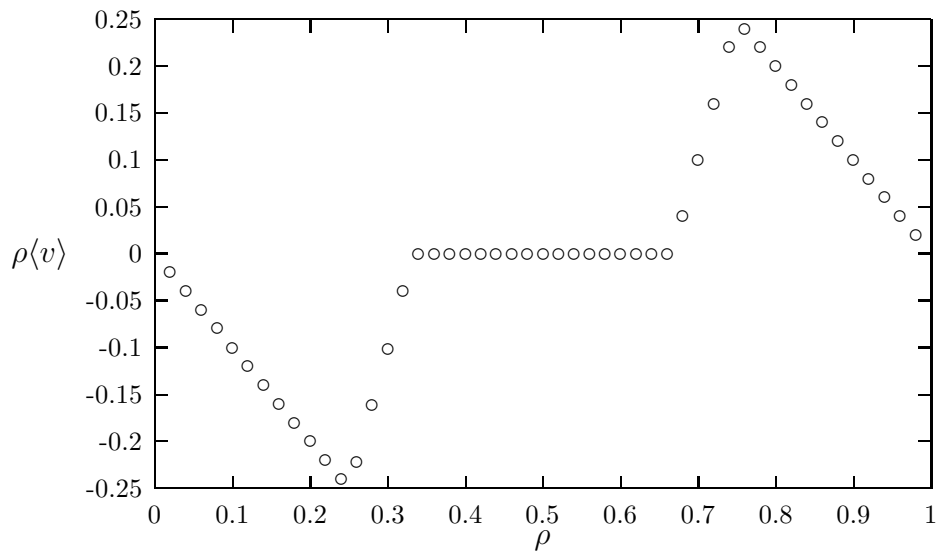

\begin{flushright}
\include{fig3}
\end{flushright}
\caption{Fundamental diagram for Rule $3167653058$. Small circles represent
numerical results.}
\settowidth{\unitlength}{\tt{0}} 
\end{figure}
\begin{figure}
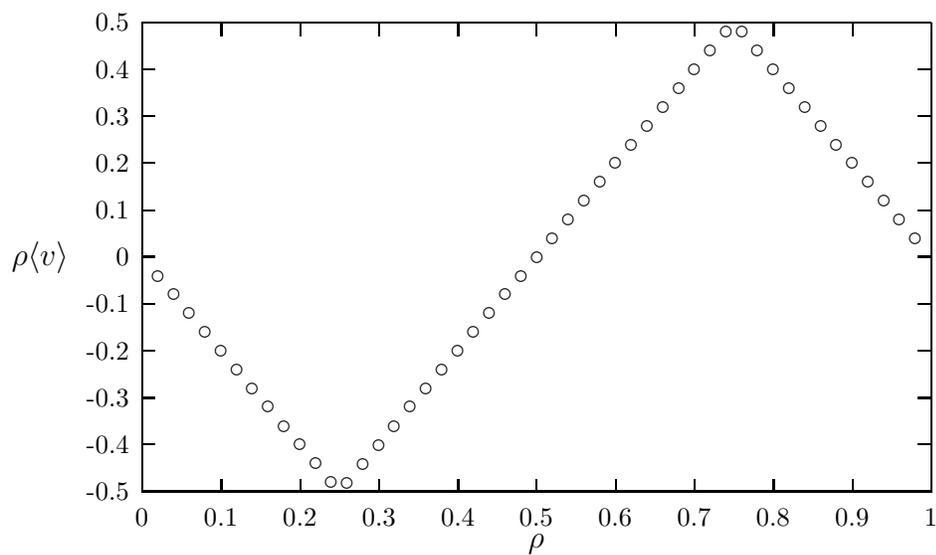

\begin{flushright}
\include{fig4}
\end{flushright}
\caption{Fundamental diagram for Rule $3221127170$. Small circles represent
numerical results.}
\settowidth{\unitlength}{\tt{0}} 
\end{figure}
\begin{figure}
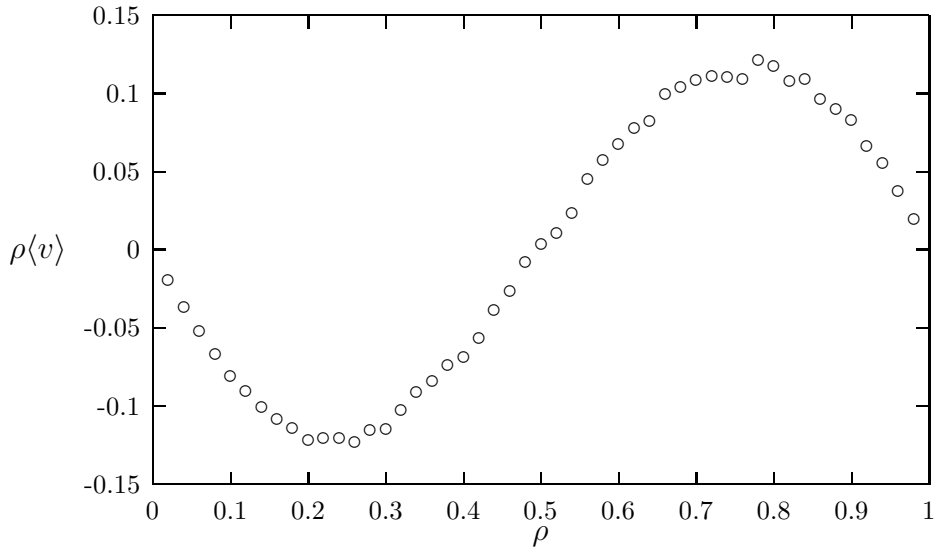

\begin{flushright}
\include{fig5}
\end{flushright}
\caption{Fundamental diagram for Rule $3707031748$. Small circles represent
numerical results.}
\settowidth{\unitlength}{\tt{0}} 
\end{figure}

The 4 remaining rules are similar to the 4-input rules 60200 and 48770
in the sense that they have similar flow diagrams and that, for
$\rho=\frac{1}{2}$, they mimic pseudo-random walkers.
These rules are:

\begin{itemize}
\item Rule 3221127170 ($r_l=2$, $r_r=2$), whose motion representation is
\[
\mbox{\jumptwoleft{0001}{4}{3}},
\mbox{\zerojump{1111}{4}{1}},
\mbox{\jumponeleft{1001}{4}{3}},
\mbox{\zerojump{101}{3}{2}},
\mbox{\jumponeright{110}{3}{1}},
\mbox{\jumponeright{1110}{4}{1}},
\]
and Rule 3937086120 obtained by reflection.
\item Rule 3707031748 ($r_l=2$, $r_r=2$), whose motion representation is
\[
\mbox{\jumponeleft{0010}{4}{2}},
\mbox{\zerojump{0011}{4}{2}},
\mbox{\zerojump{1100}{4}{1}},
\mbox{\zerojump{0111}{4}{2}},
\mbox{\zerojump{101}{3}{2}},
\mbox{\jumponeright{1101}{4}{1}},
\mbox{\zerojump{1111}{4}{2}},
\]
and Rule 416291200 obtained by reflection.
\end{itemize}
The fundamental diagrams of Rules 3221127170 and 3707031748 are
represented, respectively, in Figures 4 and 5.
Here again we verified that the corresponding stochastic processes are
not Gaussian. We have found that their Hurst exponents are equal for all
of them to $0.57\pm0.02$. We have no explanation why Rules 3221127170
and 3707031748 should have the same exponent.

\section{Approximate methods}
The mean-field approximation, which neglects correlations in space and
time, yields, for these systems, an exact but trivial result. Let
$\rho(t)$ denotes the particles density at time $t$. To find the
expression of $\rho(t+1)$ as a function of $\rho(t)$, we have to find
all the preimages of 1 by the $n$-input rule $f$. According
to~(\ref{cond-0}) all these preimages contain at least one 1.
Moreover, among all the preimages containing exactly $k+1$ times the
digit 1 ($0\le k\le n-1$), according to the conditions~(\ref{L-cond})
for $L=n$, only ${n-1}\choose k$ have a preimage equal to 1.
Therefore,
\begin{eqnarray*}
\rho(t+1) &=&\rho(t)\Bigg(\sum_{k=0}^{n-1}
{{n-1}\choose k}(\rho(t))^k(1-\rho(t))^{n-k-1}\Bigg)\\
&=&\rho(t)\big(\rho(t)+(1-\rho(t))\big)^{n-1}\\ &=&\rho(t),
\label{mfa}
\end{eqnarray*}
which expresses that density is conserved.

There exists a variety of other approximate methods which, taking into
account short-range correlations, improve the prediction of the
mean-field approximation. Instead of expressing the evolution of the
CA in terms of 1-block probabilities, they express it in terms of
$n$-block probabilities (Gutowitz {\it et al\/}, 1987). For example,
in the case of a 4-input rules, the evolution of the 2-block probability
distribution is given by
\[
P(a_1a_2) = \sum_{b_0,b_1,b_2,b_3,b_4\in\{0,1\}}
w(a_1a_2\mid b_0b_1b_2b_3b_4)P(b_0b_1b_2b_3b_4),
\]
where $P(a_1a_2)$ is the probability of block $a_1a_2$, and
\[
w(a_1a_2\mid b_0b_1b_2b_3b_4)=w(a_1\mid b_0b_1b_2b_3)w(a_2\mid b_1b_2b_3b_4)
\]
is the conditional probability that the 4-input rule maps the 5-block
$b_0b_1b_2b_3b_4$ into the 2-block $a_1a_2$. This equation is exact.
The approximation consists in replacing the 5-block probability
$P(b_0b_1b_2b_3b_4)$ in terms of 2-block probabilities. That is,
\[
P(b_0b_1b_2b_3b_4)=
\frac{P(b_0b_1)P(b_1b_2)P(b_2b_3)P(b_3b_4)}
{\Big(P(b_10)+P(b_11)\Big)\Big(P(b_20)+P(b_21)\Big)\Big(P(b_30)+P(b_31)\Big)}.
\]
We applied this method up to approximation of order 3 (mean-field being
order 1) to 4-input Rule 60200. The results are not exact, but for the flow
diagram the agreement with our numerical results is extremely good
(Figure~2).

\section{Conclusion} We have established necessary and sufficient
conditions to be satisfied by any one-dimensional cellular automaton
rule conserving the number of active sites. This result has been used
to determine all the 4- and 5-input one-dimensional cellular automaton
rules having this property. These rules express the evolution of
one-dimensional systems of interacting particles whose number is
conserved. Simple deterministic highway traffic rules belong to that
class of rules. These rules are a natural generalization of
deterministic traffic rules already studied. We have studied in more
detail (flow diagram, local structure approximation) some of our rules
allowing motion of the particles in both directions. When the particle
density is equal to $\frac{1}{2}$, these rules mimic the evolution of
pseudo-random walkers. Numerical evidence seems to indicate that the
motion of these walkers might be non-Gaussian.

\section*{References}
\begin{harvard}

\item[] Boccara N, Goles E, Mart\'{\i}nez S and Picco P eds. 1993 {\it
Cellular Automata and Cooperative Phenomena, Proc. of a  Workshop, Les
Houches} (Dordrecht: Kluwer)

\item[] Denning P, Dennis J and Qualitz J 1978 {\it Machines, Languages
and Computation} (Englewood Cliffs: Prentice Hall)

\item[] Farmer D, Toffoli T and Wolfram S eds. 1984 {\it Cellular
Automata, Proc. Interdisciplinary Workshop, Los Alamos} (Amsterdam:
North-Holland)

\item[] Feder J 1988 {\it Fractals} (New York: Plenum Press)

\item[] Fuk\'s H and Boccara N 1998 {\it Int. J. Mod. Phys. C} {\bf 9} 1

\item[] Fukui M and Ishibashi Y 1995 {\it J. Phys. Soc. Japan} {\bf 64}
1868

\item[] Gutowitz H A, Victor J D and Knight B W 1987 {\it Physica
D} {\bf 28} 18

\item[] Gutowitz H ed. 1990 {\it Cellular Automata: Theory and
Experiments, Proc. of a Workshop, Los Alamos} (Amsterdam:
North-Holland)

\item[] Hopcroft J E and Ullmam J A 1986 {\it Formal Languages and
their Relation to Automata} (Reading: Addison-Wesley)

\item[] Hurst H E 1951 {\it Transactions of the American Society of
Civil Engineering} {\bf 116} 770

\item[] Hurst H E, Black R P and Simaika Y N 1965 {\it Long-Term
Storage: An Experimental Study} (London: Constable)

\item[] Nagel K and Schreckenberg M 1992 {\it J. Physique I} {\bf 2}
{2221}

\item[] Manneville P, Boccara N, Vichniac G and Bidaux R eds. 1989 {\it
Cellular Automata and Modeling of Complex Systems, Proc. of a
Workshop, Les Houches} (Heidelberg: Springer-Verlag)

\item[] Weiss G H 1994 {\it Aspects and Applications of the Random
Walk} (Amsterdam: North-Holland)

\item[] Wolfram S 1983 {\it Rev. Mod. Phys.} {\bf 55} 601

\item[] Wolfram S 1984 {\it Commun. Math. Phys.} {\bf 96} 15

\item[] Wolfram S 1994 {\it Cellular Automata and Complexity:
Collected Papers} (Reading, Massachusetts: Addison-Wesley)

\end{harvard}

\end{document}

%% file: fig1.tex
\unitlength 0.7mm
\linethickness{0.4pt}
\begin{center}
\begin{picture}(100,55)(-50,-26)
\qbezier(0.00,0.00)(40.00,0.00)(30.00,17.30)
\qbezier(0.00,0.00)(20.00,34.60)(30.00,17.30)
\qbezier(0.00,0.00)(-40.00,0.00)(-30.00,17.30)
\qbezier(0.00,0.00)(-20.00,34.60)(-30.00,17.30)
\qbezier(0.00,0.00)(20.00,-34.60)(0.00,-34.60)
\qbezier(0.00,0.00)(-20.00,-34.60)(0.00,-34.60)
\put(0.00,0.00){\circle*{2.00}}
\put(0.00,-34.60){\circle*{2.00}}
\put(14.77,19.90){\circle*{2.00}}
\put(25.67,3.33){\circle*{2.00}}
\put(-14.77,19.90){\circle*{2.00}}
\put(-25.67,3.33){\circle*{2.00}}
\multiput(12.69,0.55)(-0.23,0.10){6}{\line(-1,0){0.23}}
\multiput(12.69,0.55)(-0.15,-0.11){8}{\line(-1,0){0.15}}
\put(7.37,11.39){\line(0,1){1.40}}
\multiput(7.37,11.39)(0.19,0.11){7}{\line(1,0){0.19}}
\multiput(-12.69,0.55)(0.23,0.10){6}{\line(1,0){0.23}}
\multiput(-12.69,0.55)(0.15,-0.11){8}{\line(1,0){0.15}}
\put(-7.37,11.39){\line(0,1){1.40}}
\multiput(-7.37,11.39)(-0.19,0.11){7}{\line(-1,0){0.19}}
\multiput(29.87,17.47)(0.25,-0.10){4}{\line(1,0){0.25}}
\put(29.87,17.47){\line(0,-1){1.04}}
\multiput(-29.87,17.47)(-0.25,-0.10){4}{\line(-1,0){0.25}}
\put(-29.87,17.47){\line(0,-1){1.04}}
\multiput(7.48,-15.04)(-0.19,0.12){8}{\line(-1,0){0.19}}
\multiput(7.48,-15.04)(0.07,0.77){2}{\line(0,1){0.77}}
\multiput(-6.87,-13.39)(-0.17,-0.12){8}{\line(-1,0){0.17}}
\multiput(-6.80,-13.39)(0.11,-0.71){2}{\line(0,-1){0.71}}
\put(10.50,-14.17){\makebox(0,0)[cc]{1}}
\put(-10.83,-14.17){\makebox(0,0)[cc]{0}}
\put(12.67,-3.00){\makebox(0,0)[cc]{1}}
\put(33.33,20.83){\makebox(0,0)[cc]{1}}
\put(3.00,15.17){\makebox(0,0)[cc]{0}}
\put(-11.83,-3.00){\makebox(0,0)[cc]{1}}
\put(-35.00,20.33){\makebox(0,0)[cc]{0}}
\put(-4.67,15.17){\makebox(0,0)[cc]{0}}
\end{picture}
\end{center}

%% file: fig2.tex
\setlength{\unitlength}{0.240900pt}
\ifx\plotpoint\undefined\newsavebox{\plotpoint}\fi
\sbox{\plotpoint}{\rule[-0.200pt]{0.400pt}{0.400pt}}%
\begin{picture}(1500,900)(0,0)
\font\gnuplot=cmr10 at 10pt
\gnuplot
\sbox{\plotpoint}{\rule[-0.200pt]{0.400pt}{0.400pt}}%
\put(201.0,122.0){\rule[-0.200pt]{4.818pt}{0.400pt}}
\put(181,122){\makebox(0,0)[r]{-0.4}}
\put(1420.0,122.0){\rule[-0.200pt]{4.818pt}{0.400pt}}
\put(201.0,214.0){\rule[-0.200pt]{4.818pt}{0.400pt}}
\put(181,214){\makebox(0,0)[r]{-0.3}}
\put(1420.0,214.0){\rule[-0.200pt]{4.818pt}{0.400pt}}
\put(201.0,306.0){\rule[-0.200pt]{4.818pt}{0.400pt}}
\put(181,306){\makebox(0,0)[r]{-0.2}}
\put(1420.0,306.0){\rule[-0.200pt]{4.818pt}{0.400pt}}
\put(201.0,398.0){\rule[-0.200pt]{4.818pt}{0.400pt}}
\put(181,398){\makebox(0,0)[r]{-0.1}}
\put(1420.0,398.0){\rule[-0.200pt]{4.818pt}{0.400pt}}
\put(201.0,490.0){\rule[-0.200pt]{4.818pt}{0.400pt}}
\put(181,490){\makebox(0,0)[r]{0}}
\put(1420.0,490.0){\rule[-0.200pt]{4.818pt}{0.400pt}}
\put(201.0,583.0){\rule[-0.200pt]{4.818pt}{0.400pt}}
\put(181,583){\makebox(0,0)[r]{0.1}}
\put(1420.0,583.0){\rule[-0.200pt]{4.818pt}{0.400pt}}
\put(201.0,675.0){\rule[-0.200pt]{4.818pt}{0.400pt}}
\put(181,675){\makebox(0,0)[r]{0.2}}
\put(1420.0,675.0){\rule[-0.200pt]{4.818pt}{0.400pt}}
\put(201.0,767.0){\rule[-0.200pt]{4.818pt}{0.400pt}}
\put(181,767){\makebox(0,0)[r]{0.3}}
\put(1420.0,767.0){\rule[-0.200pt]{4.818pt}{0.400pt}}
\put(201.0,859.0){\rule[-0.200pt]{4.818pt}{0.400pt}}
\put(181,859){\makebox(0,0)[r]{0.4}}
\put(1420.0,859.0){\rule[-0.200pt]{4.818pt}{0.400pt}}
\put(201.0,122.0){\rule[-0.200pt]{0.400pt}{4.818pt}}
\put(201,81){\makebox(0,0){0}}
\put(201.0,839.0){\rule[-0.200pt]{0.400pt}{4.818pt}}
\put(449.0,122.0){\rule[-0.200pt]{0.400pt}{4.818pt}}
\put(449,81){\makebox(0,0){0.2}}
\put(449.0,839.0){\rule[-0.200pt]{0.400pt}{4.818pt}}
\put(697.0,122.0){\rule[-0.200pt]{0.400pt}{4.818pt}}
\put(697,81){\makebox(0,0){0.4}}
\put(697.0,839.0){\rule[-0.200pt]{0.400pt}{4.818pt}}
\put(944.0,122.0){\rule[-0.200pt]{0.400pt}{4.818pt}}
\put(944,81){\makebox(0,0){0.6}}
\put(944.0,839.0){\rule[-0.200pt]{0.400pt}{4.818pt}}
\put(1192.0,122.0){\rule[-0.200pt]{0.400pt}{4.818pt}}
\put(1192,81){\makebox(0,0){0.8}}
\put(1192.0,839.0){\rule[-0.200pt]{0.400pt}{4.818pt}}
\put(1440.0,122.0){\rule[-0.200pt]{0.400pt}{4.818pt}}
\put(1440,81){\makebox(0,0){1}}
\put(1440.0,839.0){\rule[-0.200pt]{0.400pt}{4.818pt}}
\put(201.0,122.0){\rule[-0.200pt]{298.475pt}{0.400pt}}
\put(1440.0,122.0){\rule[-0.200pt]{0.400pt}{177.543pt}}
\put(201.0,859.0){\rule[-0.200pt]{298.475pt}{0.400pt}}
\put(41,490){\makebox(0,0){$\rho \langle v \rangle$}}
\put(820,40){\makebox(0,0){$\rho$}}
\put(201.0,122.0){\rule[-0.200pt]{0.400pt}{177.543pt}}
\put(201,491){\circle{18}}
\put(226,509){\circle{18}}
\put(251,527){\circle{18}}
\put(275,546){\circle{18}}
\put(300,564){\circle{18}}
\put(325,583){\circle{18}}
\put(350,601){\circle{18}}
\put(374,619){\circle{18}}
\put(399,638){\circle{18}}
\put(424,656){\circle{18}}
\put(449,675){\circle{18}}
\put(474,693){\circle{18}}
\put(498,712){\circle{18}}
\put(523,730){\circle{18}}
\put(548,748){\circle{18}}
\put(573,767){\circle{18}}
\put(597,785){\circle{18}}
\put(622,785){\circle{18}}
\put(647,748){\circle{18}}
\put(672,712){\circle{18}}
\put(697,675){\circle{18}}
\put(721,638){\circle{18}}
\put(746,601){\circle{18}}
\put(771,564){\circle{18}}
\put(796,527){\circle{18}}
\put(821,491){\circle{18}}
\put(845,454){\circle{18}}
\put(870,417){\circle{18}}
\put(895,380){\circle{18}}
\put(920,343){\circle{18}}
\put(944,306){\circle{18}}
\put(969,269){\circle{18}}
\put(994,233){\circle{18}}
\put(1019,196){\circle{18}}
\put(1044,196){\circle{18}}
\put(1068,214){\circle{18}}
\put(1093,233){\circle{18}}
\put(1118,251){\circle{18}}
\put(1143,269){\circle{18}}
\put(1167,288){\circle{18}}
\put(1192,306){\circle{18}}
\put(1217,325){\circle{18}}
\put(1242,343){\circle{18}}
\put(1267,362){\circle{18}}
\put(1291,380){\circle{18}}
\put(1316,398){\circle{18}}
\put(1341,417){\circle{18}}
\put(1366,435){\circle{18}}
\put(1390,454){\circle{18}}
\put(1415,472){\circle{18}}
\put(1440,491){\circle{18}}
\put(201,491){\usebox{\plotpoint}}
\multiput(201.00,491.59)(0.728,0.489){15}{\rule{0.678pt}{0.118pt}}
\multiput(201.00,490.17)(11.593,9.000){2}{\rule{0.339pt}{0.400pt}}
\multiput(214.00,500.59)(0.669,0.489){15}{\rule{0.633pt}{0.118pt}}
\multiput(214.00,499.17)(10.685,9.000){2}{\rule{0.317pt}{0.400pt}}
\multiput(226.00,509.59)(0.728,0.489){15}{\rule{0.678pt}{0.118pt}}
\multiput(226.00,508.17)(11.593,9.000){2}{\rule{0.339pt}{0.400pt}}
\multiput(239.00,518.58)(0.600,0.491){17}{\rule{0.580pt}{0.118pt}}
\multiput(239.00,517.17)(10.796,10.000){2}{\rule{0.290pt}{0.400pt}}
\multiput(251.00,528.59)(0.728,0.489){15}{\rule{0.678pt}{0.118pt}}
\multiput(251.00,527.17)(11.593,9.000){2}{\rule{0.339pt}{0.400pt}}
\multiput(264.00,537.59)(0.669,0.489){15}{\rule{0.633pt}{0.118pt}}
\multiput(264.00,536.17)(10.685,9.000){2}{\rule{0.317pt}{0.400pt}}
\multiput(276.00,546.58)(0.652,0.491){17}{\rule{0.620pt}{0.118pt}}
\multiput(276.00,545.17)(11.713,10.000){2}{\rule{0.310pt}{0.400pt}}
\multiput(289.00,556.59)(0.669,0.489){15}{\rule{0.633pt}{0.118pt}}
\multiput(289.00,555.17)(10.685,9.000){2}{\rule{0.317pt}{0.400pt}}
\multiput(301.00,565.59)(0.728,0.489){15}{\rule{0.678pt}{0.118pt}}
\multiput(301.00,564.17)(11.593,9.000){2}{\rule{0.339pt}{0.400pt}}
\multiput(314.00,574.58)(0.600,0.491){17}{\rule{0.580pt}{0.118pt}}
\multiput(314.00,573.17)(10.796,10.000){2}{\rule{0.290pt}{0.400pt}}
\multiput(326.00,584.59)(0.728,0.489){15}{\rule{0.678pt}{0.118pt}}
\multiput(326.00,583.17)(11.593,9.000){2}{\rule{0.339pt}{0.400pt}}
\multiput(339.00,593.59)(0.669,0.489){15}{\rule{0.633pt}{0.118pt}}
\multiput(339.00,592.17)(10.685,9.000){2}{\rule{0.317pt}{0.400pt}}
\multiput(351.00,602.59)(0.728,0.489){15}{\rule{0.678pt}{0.118pt}}
\multiput(351.00,601.17)(11.593,9.000){2}{\rule{0.339pt}{0.400pt}}
\multiput(364.00,611.58)(0.600,0.491){17}{\rule{0.580pt}{0.118pt}}
\multiput(364.00,610.17)(10.796,10.000){2}{\rule{0.290pt}{0.400pt}}
\multiput(376.00,621.59)(0.728,0.489){15}{\rule{0.678pt}{0.118pt}}
\multiput(376.00,620.17)(11.593,9.000){2}{\rule{0.339pt}{0.400pt}}
\multiput(389.00,630.59)(0.669,0.489){15}{\rule{0.633pt}{0.118pt}}
\multiput(389.00,629.17)(10.685,9.000){2}{\rule{0.317pt}{0.400pt}}
\multiput(401.00,639.58)(0.652,0.491){17}{\rule{0.620pt}{0.118pt}}
\multiput(401.00,638.17)(11.713,10.000){2}{\rule{0.310pt}{0.400pt}}
\multiput(414.00,649.59)(0.669,0.489){15}{\rule{0.633pt}{0.118pt}}
\multiput(414.00,648.17)(10.685,9.000){2}{\rule{0.317pt}{0.400pt}}
\multiput(426.00,658.59)(0.728,0.489){15}{\rule{0.678pt}{0.118pt}}
\multiput(426.00,657.17)(11.593,9.000){2}{\rule{0.339pt}{0.400pt}}
\multiput(439.00,667.58)(0.600,0.491){17}{\rule{0.580pt}{0.118pt}}
\multiput(439.00,666.17)(10.796,10.000){2}{\rule{0.290pt}{0.400pt}}
\multiput(451.00,677.59)(0.728,0.489){15}{\rule{0.678pt}{0.118pt}}
\multiput(451.00,676.17)(11.593,9.000){2}{\rule{0.339pt}{0.400pt}}
\multiput(464.00,686.59)(0.669,0.489){15}{\rule{0.633pt}{0.118pt}}
\multiput(464.00,685.17)(10.685,9.000){2}{\rule{0.317pt}{0.400pt}}
\multiput(476.00,695.58)(0.652,0.491){17}{\rule{0.620pt}{0.118pt}}
\multiput(476.00,694.17)(11.713,10.000){2}{\rule{0.310pt}{0.400pt}}
\multiput(489.00,705.59)(0.669,0.489){15}{\rule{0.633pt}{0.118pt}}
\multiput(489.00,704.17)(10.685,9.000){2}{\rule{0.317pt}{0.400pt}}
\multiput(501.00,714.59)(0.728,0.489){15}{\rule{0.678pt}{0.118pt}}
\multiput(501.00,713.17)(11.593,9.000){2}{\rule{0.339pt}{0.400pt}}
\multiput(514.00,723.59)(0.669,0.489){15}{\rule{0.633pt}{0.118pt}}
\multiput(514.00,722.17)(10.685,9.000){2}{\rule{0.317pt}{0.400pt}}
\multiput(526.00,732.58)(0.652,0.491){17}{\rule{0.620pt}{0.118pt}}
\multiput(526.00,731.17)(11.713,10.000){2}{\rule{0.310pt}{0.400pt}}
\multiput(539.00,742.59)(0.669,0.489){15}{\rule{0.633pt}{0.118pt}}
\multiput(539.00,741.17)(10.685,9.000){2}{\rule{0.317pt}{0.400pt}}
\multiput(551.00,751.59)(0.728,0.489){15}{\rule{0.678pt}{0.118pt}}
\multiput(551.00,750.17)(11.593,9.000){2}{\rule{0.339pt}{0.400pt}}
\multiput(564.00,760.58)(0.600,0.491){17}{\rule{0.580pt}{0.118pt}}
\multiput(564.00,759.17)(10.796,10.000){2}{\rule{0.290pt}{0.400pt}}
\multiput(576.00,770.59)(0.728,0.489){15}{\rule{0.678pt}{0.118pt}}
\multiput(576.00,769.17)(11.593,9.000){2}{\rule{0.339pt}{0.400pt}}
\multiput(589.00,779.59)(0.669,0.489){15}{\rule{0.633pt}{0.118pt}}
\multiput(589.00,778.17)(10.685,9.000){2}{\rule{0.317pt}{0.400pt}}
\multiput(601.00,788.58)(0.652,0.491){17}{\rule{0.620pt}{0.118pt}}
\multiput(601.00,787.17)(11.713,10.000){2}{\rule{0.310pt}{0.400pt}}
\multiput(614.58,795.16)(0.493,-0.734){23}{\rule{0.119pt}{0.685pt}}
\multiput(613.17,796.58)(13.000,-17.579){2}{\rule{0.400pt}{0.342pt}}
\multiput(627.58,775.96)(0.492,-0.798){21}{\rule{0.119pt}{0.733pt}}
\multiput(626.17,777.48)(12.000,-17.478){2}{\rule{0.400pt}{0.367pt}}
\multiput(639.58,757.29)(0.493,-0.695){23}{\rule{0.119pt}{0.654pt}}
\multiput(638.17,758.64)(13.000,-16.643){2}{\rule{0.400pt}{0.327pt}}
\multiput(652.58,738.96)(0.492,-0.798){21}{\rule{0.119pt}{0.733pt}}
\multiput(651.17,740.48)(12.000,-17.478){2}{\rule{0.400pt}{0.367pt}}
\multiput(664.58,720.29)(0.493,-0.695){23}{\rule{0.119pt}{0.654pt}}
\multiput(663.17,721.64)(13.000,-16.643){2}{\rule{0.400pt}{0.327pt}}
\multiput(677.58,701.96)(0.492,-0.798){21}{\rule{0.119pt}{0.733pt}}
\multiput(676.17,703.48)(12.000,-17.478){2}{\rule{0.400pt}{0.367pt}}
\multiput(689.58,683.16)(0.493,-0.734){23}{\rule{0.119pt}{0.685pt}}
\multiput(688.17,684.58)(13.000,-17.579){2}{\rule{0.400pt}{0.342pt}}
\multiput(702.58,664.09)(0.492,-0.755){21}{\rule{0.119pt}{0.700pt}}
\multiput(701.17,665.55)(12.000,-16.547){2}{\rule{0.400pt}{0.350pt}}
\multiput(714.58,646.16)(0.493,-0.734){23}{\rule{0.119pt}{0.685pt}}
\multiput(713.17,647.58)(13.000,-17.579){2}{\rule{0.400pt}{0.342pt}}
\multiput(727.58,626.96)(0.492,-0.798){21}{\rule{0.119pt}{0.733pt}}
\multiput(726.17,628.48)(12.000,-17.478){2}{\rule{0.400pt}{0.367pt}}
\multiput(739.58,608.29)(0.493,-0.695){23}{\rule{0.119pt}{0.654pt}}
\multiput(738.17,609.64)(13.000,-16.643){2}{\rule{0.400pt}{0.327pt}}
\multiput(752.58,589.96)(0.492,-0.798){21}{\rule{0.119pt}{0.733pt}}
\multiput(751.17,591.48)(12.000,-17.478){2}{\rule{0.400pt}{0.367pt}}
\multiput(764.58,571.29)(0.493,-0.695){23}{\rule{0.119pt}{0.654pt}}
\multiput(763.17,572.64)(13.000,-16.643){2}{\rule{0.400pt}{0.327pt}}
\multiput(777.58,552.96)(0.492,-0.798){21}{\rule{0.119pt}{0.733pt}}
\multiput(776.17,554.48)(12.000,-17.478){2}{\rule{0.400pt}{0.367pt}}
\multiput(789.58,534.16)(0.493,-0.734){23}{\rule{0.119pt}{0.685pt}}
\multiput(788.17,535.58)(13.000,-17.579){2}{\rule{0.400pt}{0.342pt}}
\multiput(802.58,515.09)(0.492,-0.755){21}{\rule{0.119pt}{0.700pt}}
\multiput(801.17,516.55)(12.000,-16.547){2}{\rule{0.400pt}{0.350pt}}
\multiput(814.58,497.16)(0.493,-0.734){23}{\rule{0.119pt}{0.685pt}}
\multiput(813.17,498.58)(13.000,-17.579){2}{\rule{0.400pt}{0.342pt}}
\multiput(827.58,478.09)(0.492,-0.755){21}{\rule{0.119pt}{0.700pt}}
\multiput(826.17,479.55)(12.000,-16.547){2}{\rule{0.400pt}{0.350pt}}
\multiput(839.58,460.16)(0.493,-0.734){23}{\rule{0.119pt}{0.685pt}}
\multiput(838.17,461.58)(13.000,-17.579){2}{\rule{0.400pt}{0.342pt}}
\multiput(852.58,440.96)(0.492,-0.798){21}{\rule{0.119pt}{0.733pt}}
\multiput(851.17,442.48)(12.000,-17.478){2}{\rule{0.400pt}{0.367pt}}
\multiput(864.58,422.29)(0.493,-0.695){23}{\rule{0.119pt}{0.654pt}}
\multiput(863.17,423.64)(13.000,-16.643){2}{\rule{0.400pt}{0.327pt}}
\multiput(877.58,403.96)(0.492,-0.798){21}{\rule{0.119pt}{0.733pt}}
\multiput(876.17,405.48)(12.000,-17.478){2}{\rule{0.400pt}{0.367pt}}
\multiput(889.58,385.29)(0.493,-0.695){23}{\rule{0.119pt}{0.654pt}}
\multiput(888.17,386.64)(13.000,-16.643){2}{\rule{0.400pt}{0.327pt}}
\multiput(902.58,366.96)(0.492,-0.798){21}{\rule{0.119pt}{0.733pt}}
\multiput(901.17,368.48)(12.000,-17.478){2}{\rule{0.400pt}{0.367pt}}
\multiput(914.58,348.16)(0.493,-0.734){23}{\rule{0.119pt}{0.685pt}}
\multiput(913.17,349.58)(13.000,-17.579){2}{\rule{0.400pt}{0.342pt}}
\multiput(927.58,329.09)(0.492,-0.755){21}{\rule{0.119pt}{0.700pt}}
\multiput(926.17,330.55)(12.000,-16.547){2}{\rule{0.400pt}{0.350pt}}
\multiput(939.58,311.16)(0.493,-0.734){23}{\rule{0.119pt}{0.685pt}}
\multiput(938.17,312.58)(13.000,-17.579){2}{\rule{0.400pt}{0.342pt}}
\multiput(952.58,291.96)(0.492,-0.798){21}{\rule{0.119pt}{0.733pt}}
\multiput(951.17,293.48)(12.000,-17.478){2}{\rule{0.400pt}{0.367pt}}
\multiput(964.58,273.29)(0.493,-0.695){23}{\rule{0.119pt}{0.654pt}}
\multiput(963.17,274.64)(13.000,-16.643){2}{\rule{0.400pt}{0.327pt}}
\multiput(977.58,254.96)(0.492,-0.798){21}{\rule{0.119pt}{0.733pt}}
\multiput(976.17,256.48)(12.000,-17.478){2}{\rule{0.400pt}{0.367pt}}
\multiput(989.58,236.29)(0.493,-0.695){23}{\rule{0.119pt}{0.654pt}}
\multiput(988.17,237.64)(13.000,-16.643){2}{\rule{0.400pt}{0.327pt}}
\multiput(1002.58,217.96)(0.492,-0.798){21}{\rule{0.119pt}{0.733pt}}
\multiput(1001.17,219.48)(12.000,-17.478){2}{\rule{0.400pt}{0.367pt}}
\multiput(1014.58,199.16)(0.493,-0.734){23}{\rule{0.119pt}{0.685pt}}
\multiput(1013.17,200.58)(13.000,-17.579){2}{\rule{0.400pt}{0.342pt}}
\multiput(1027.00,183.58)(0.652,0.491){17}{\rule{0.620pt}{0.118pt}}
\multiput(1027.00,182.17)(11.713,10.000){2}{\rule{0.310pt}{0.400pt}}
\multiput(1040.00,193.59)(0.669,0.489){15}{\rule{0.633pt}{0.118pt}}
\multiput(1040.00,192.17)(10.685,9.000){2}{\rule{0.317pt}{0.400pt}}
\multiput(1052.00,202.59)(0.728,0.489){15}{\rule{0.678pt}{0.118pt}}
\multiput(1052.00,201.17)(11.593,9.000){2}{\rule{0.339pt}{0.400pt}}
\multiput(1065.00,211.58)(0.600,0.491){17}{\rule{0.580pt}{0.118pt}}
\multiput(1065.00,210.17)(10.796,10.000){2}{\rule{0.290pt}{0.400pt}}
\multiput(1077.00,221.59)(0.728,0.489){15}{\rule{0.678pt}{0.118pt}}
\multiput(1077.00,220.17)(11.593,9.000){2}{\rule{0.339pt}{0.400pt}}
\multiput(1090.00,230.59)(0.669,0.489){15}{\rule{0.633pt}{0.118pt}}
\multiput(1090.00,229.17)(10.685,9.000){2}{\rule{0.317pt}{0.400pt}}
\multiput(1102.00,239.58)(0.652,0.491){17}{\rule{0.620pt}{0.118pt}}
\multiput(1102.00,238.17)(11.713,10.000){2}{\rule{0.310pt}{0.400pt}}
\multiput(1115.00,249.59)(0.669,0.489){15}{\rule{0.633pt}{0.118pt}}
\multiput(1115.00,248.17)(10.685,9.000){2}{\rule{0.317pt}{0.400pt}}
\multiput(1127.00,258.59)(0.728,0.489){15}{\rule{0.678pt}{0.118pt}}
\multiput(1127.00,257.17)(11.593,9.000){2}{\rule{0.339pt}{0.400pt}}
\multiput(1140.00,267.59)(0.669,0.489){15}{\rule{0.633pt}{0.118pt}}
\multiput(1140.00,266.17)(10.685,9.000){2}{\rule{0.317pt}{0.400pt}}
\multiput(1152.00,276.58)(0.652,0.491){17}{\rule{0.620pt}{0.118pt}}
\multiput(1152.00,275.17)(11.713,10.000){2}{\rule{0.310pt}{0.400pt}}
\multiput(1165.00,286.59)(0.669,0.489){15}{\rule{0.633pt}{0.118pt}}
\multiput(1165.00,285.17)(10.685,9.000){2}{\rule{0.317pt}{0.400pt}}
\multiput(1177.00,295.59)(0.728,0.489){15}{\rule{0.678pt}{0.118pt}}
\multiput(1177.00,294.17)(11.593,9.000){2}{\rule{0.339pt}{0.400pt}}
\multiput(1190.00,304.58)(0.600,0.491){17}{\rule{0.580pt}{0.118pt}}
\multiput(1190.00,303.17)(10.796,10.000){2}{\rule{0.290pt}{0.400pt}}
\multiput(1202.00,314.59)(0.728,0.489){15}{\rule{0.678pt}{0.118pt}}
\multiput(1202.00,313.17)(11.593,9.000){2}{\rule{0.339pt}{0.400pt}}
\multiput(1215.00,323.59)(0.669,0.489){15}{\rule{0.633pt}{0.118pt}}
\multiput(1215.00,322.17)(10.685,9.000){2}{\rule{0.317pt}{0.400pt}}
\multiput(1227.00,332.58)(0.652,0.491){17}{\rule{0.620pt}{0.118pt}}
\multiput(1227.00,331.17)(11.713,10.000){2}{\rule{0.310pt}{0.400pt}}
\multiput(1240.00,342.59)(0.669,0.489){15}{\rule{0.633pt}{0.118pt}}
\multiput(1240.00,341.17)(10.685,9.000){2}{\rule{0.317pt}{0.400pt}}
\multiput(1252.00,351.59)(0.728,0.489){15}{\rule{0.678pt}{0.118pt}}
\multiput(1252.00,350.17)(11.593,9.000){2}{\rule{0.339pt}{0.400pt}}
\multiput(1265.00,360.58)(0.600,0.491){17}{\rule{0.580pt}{0.118pt}}
\multiput(1265.00,359.17)(10.796,10.000){2}{\rule{0.290pt}{0.400pt}}
\multiput(1277.00,370.59)(0.728,0.489){15}{\rule{0.678pt}{0.118pt}}
\multiput(1277.00,369.17)(11.593,9.000){2}{\rule{0.339pt}{0.400pt}}
\multiput(1290.00,379.59)(0.669,0.489){15}{\rule{0.633pt}{0.118pt}}
\multiput(1290.00,378.17)(10.685,9.000){2}{\rule{0.317pt}{0.400pt}}
\multiput(1302.00,388.59)(0.728,0.489){15}{\rule{0.678pt}{0.118pt}}
\multiput(1302.00,387.17)(11.593,9.000){2}{\rule{0.339pt}{0.400pt}}
\multiput(1315.00,397.58)(0.600,0.491){17}{\rule{0.580pt}{0.118pt}}
\multiput(1315.00,396.17)(10.796,10.000){2}{\rule{0.290pt}{0.400pt}}
\multiput(1327.00,407.59)(0.728,0.489){15}{\rule{0.678pt}{0.118pt}}
\multiput(1327.00,406.17)(11.593,9.000){2}{\rule{0.339pt}{0.400pt}}
\multiput(1340.00,416.59)(0.669,0.489){15}{\rule{0.633pt}{0.118pt}}
\multiput(1340.00,415.17)(10.685,9.000){2}{\rule{0.317pt}{0.400pt}}
\multiput(1352.00,425.58)(0.652,0.491){17}{\rule{0.620pt}{0.118pt}}
\multiput(1352.00,424.17)(11.713,10.000){2}{\rule{0.310pt}{0.400pt}}
\multiput(1365.00,435.59)(0.669,0.489){15}{\rule{0.633pt}{0.118pt}}
\multiput(1365.00,434.17)(10.685,9.000){2}{\rule{0.317pt}{0.400pt}}
\multiput(1377.00,444.59)(0.728,0.489){15}{\rule{0.678pt}{0.118pt}}
\multiput(1377.00,443.17)(11.593,9.000){2}{\rule{0.339pt}{0.400pt}}
\multiput(1390.00,453.58)(0.600,0.491){17}{\rule{0.580pt}{0.118pt}}
\multiput(1390.00,452.17)(10.796,10.000){2}{\rule{0.290pt}{0.400pt}}
\multiput(1402.00,463.59)(0.728,0.489){15}{\rule{0.678pt}{0.118pt}}
\multiput(1402.00,462.17)(11.593,9.000){2}{\rule{0.339pt}{0.400pt}}
\multiput(1415.00,472.59)(0.669,0.489){15}{\rule{0.633pt}{0.118pt}}
\multiput(1415.00,471.17)(10.685,9.000){2}{\rule{0.317pt}{0.400pt}}
\multiput(1427.00,481.58)(0.652,0.491){17}{\rule{0.620pt}{0.118pt}}
\multiput(1427.00,480.17)(11.713,10.000){2}{\rule{0.310pt}{0.400pt}}
\end{picture}

%% file: fig3.tex
\setlength{\unitlength}{0.240900pt}
\ifx\plotpoint\undefined\newsavebox{\plotpoint}\fi
\sbox{\plotpoint}{\rule[-0.200pt]{0.400pt}{0.400pt}}%
\begin{picture}(1500,900)(0,0)
\font\gnuplot=cmr10 at 10pt
\gnuplot
\sbox{\plotpoint}{\rule[-0.200pt]{0.400pt}{0.400pt}}%
\put(221.0,122.0){\rule[-0.200pt]{4.818pt}{0.400pt}}
\put(201,122){\makebox(0,0)[r]{-0.25}}
\put(1420.0,122.0){\rule[-0.200pt]{4.818pt}{0.400pt}}
\put(221.0,196.0){\rule[-0.200pt]{4.818pt}{0.400pt}}
\put(201,196){\makebox(0,0)[r]{-0.2}}
\put(1420.0,196.0){\rule[-0.200pt]{4.818pt}{0.400pt}}
\put(221.0,269.0){\rule[-0.200pt]{4.818pt}{0.400pt}}
\put(201,269){\makebox(0,0)[r]{-0.15}}
\put(1420.0,269.0){\rule[-0.200pt]{4.818pt}{0.400pt}}
\put(221.0,343.0){\rule[-0.200pt]{4.818pt}{0.400pt}}
\put(201,343){\makebox(0,0)[r]{-0.1}}
\put(1420.0,343.0){\rule[-0.200pt]{4.818pt}{0.400pt}}
\put(221.0,417.0){\rule[-0.200pt]{4.818pt}{0.400pt}}
\put(201,417){\makebox(0,0)[r]{-0.05}}
\put(1420.0,417.0){\rule[-0.200pt]{4.818pt}{0.400pt}}
\put(221.0,491.0){\rule[-0.200pt]{4.818pt}{0.400pt}}
\put(201,491){\makebox(0,0)[r]{0}}
\put(1420.0,491.0){\rule[-0.200pt]{4.818pt}{0.400pt}}
\put(221.0,564.0){\rule[-0.200pt]{4.818pt}{0.400pt}}
\put(201,564){\makebox(0,0)[r]{0.05}}
\put(1420.0,564.0){\rule[-0.200pt]{4.818pt}{0.400pt}}
\put(221.0,638.0){\rule[-0.200pt]{4.818pt}{0.400pt}}
\put(201,638){\makebox(0,0)[r]{0.1}}
\put(1420.0,638.0){\rule[-0.200pt]{4.818pt}{0.400pt}}
\put(221.0,712.0){\rule[-0.200pt]{4.818pt}{0.400pt}}
\put(201,712){\makebox(0,0)[r]{0.15}}
\put(1420.0,712.0){\rule[-0.200pt]{4.818pt}{0.400pt}}
\put(221.0,785.0){\rule[-0.200pt]{4.818pt}{0.400pt}}
\put(201,785){\makebox(0,0)[r]{0.2}}
\put(1420.0,785.0){\rule[-0.200pt]{4.818pt}{0.400pt}}
\put(221.0,859.0){\rule[-0.200pt]{4.818pt}{0.400pt}}
\put(201,859){\makebox(0,0)[r]{0.25}}
\put(1420.0,859.0){\rule[-0.200pt]{4.818pt}{0.400pt}}
\put(221.0,122.0){\rule[-0.200pt]{0.400pt}{4.818pt}}
\put(221,81){\makebox(0,0){0}}
\put(221.0,839.0){\rule[-0.200pt]{0.400pt}{4.818pt}}
\put(343.0,122.0){\rule[-0.200pt]{0.400pt}{4.818pt}}
\put(343,81){\makebox(0,0){0.1}}
\put(343.0,839.0){\rule[-0.200pt]{0.400pt}{4.818pt}}
\put(465.0,122.0){\rule[-0.200pt]{0.400pt}{4.818pt}}
\put(465,81){\makebox(0,0){0.2}}
\put(465.0,839.0){\rule[-0.200pt]{0.400pt}{4.818pt}}
\put(587.0,122.0){\rule[-0.200pt]{0.400pt}{4.818pt}}
\put(587,81){\makebox(0,0){0.3}}
\put(587.0,839.0){\rule[-0.200pt]{0.400pt}{4.818pt}}
\put(709.0,122.0){\rule[-0.200pt]{0.400pt}{4.818pt}}
\put(709,81){\makebox(0,0){0.4}}
\put(709.0,839.0){\rule[-0.200pt]{0.400pt}{4.818pt}}
\put(830.0,122.0){\rule[-0.200pt]{0.400pt}{4.818pt}}
\put(830,81){\makebox(0,0){0.5}}
\put(830.0,839.0){\rule[-0.200pt]{0.400pt}{4.818pt}}
\put(952.0,122.0){\rule[-0.200pt]{0.400pt}{4.818pt}}
\put(952,81){\makebox(0,0){0.6}}
\put(952.0,839.0){\rule[-0.200pt]{0.400pt}{4.818pt}}
\put(1074.0,122.0){\rule[-0.200pt]{0.400pt}{4.818pt}}
\put(1074,81){\makebox(0,0){0.7}}
\put(1074.0,839.0){\rule[-0.200pt]{0.400pt}{4.818pt}}
\put(1196.0,122.0){\rule[-0.200pt]{0.400pt}{4.818pt}}
\put(1196,81){\makebox(0,0){0.8}}
\put(1196.0,839.0){\rule[-0.200pt]{0.400pt}{4.818pt}}
\put(1318.0,122.0){\rule[-0.200pt]{0.400pt}{4.818pt}}
\put(1318,81){\makebox(0,0){0.9}}
\put(1318.0,839.0){\rule[-0.200pt]{0.400pt}{4.818pt}}
\put(1440.0,122.0){\rule[-0.200pt]{0.400pt}{4.818pt}}
\put(1440,81){\makebox(0,0){1}}
\put(1440.0,839.0){\rule[-0.200pt]{0.400pt}{4.818pt}}
\put(221.0,122.0){\rule[-0.200pt]{293.657pt}{0.400pt}}
\put(1440.0,122.0){\rule[-0.200pt]{0.400pt}{177.543pt}}
\put(221.0,859.0){\rule[-0.200pt]{293.657pt}{0.400pt}}
\put(41,490){\makebox(0,0){${\rho \langle v \rangle}$}}
\put(830,40){\makebox(0,0){${\rho}$}}
\put(221.0,122.0){\rule[-0.200pt]{0.400pt}{177.543pt}}
\put(245,461){\circle{18}}
\put(270,432){\circle{18}}
\put(294,402){\circle{18}}
\put(319,373){\circle{18}}
\put(343,343){\circle{18}}
\put(367,314){\circle{18}}
\put(392,284){\circle{18}}
\put(416,255){\circle{18}}
\put(440,225){\circle{18}}
\put(465,196){\circle{18}}
\put(489,166){\circle{18}}
\put(514,137){\circle{18}}
\put(538,163){\circle{18}}
\put(562,252){\circle{18}}
\put(587,341){\circle{18}}
\put(611,432){\circle{18}}
\put(635,490){\circle{18}}
\put(660,490){\circle{18}}
\put(684,490){\circle{18}}
\put(709,490){\circle{18}}
\put(733,490){\circle{18}}
\put(757,490){\circle{18}}
\put(782,490){\circle{18}}
\put(806,490){\circle{18}}
\put(831,490){\circle{18}}
\put(855,490){\circle{18}}
\put(879,490){\circle{18}}
\put(904,490){\circle{18}}
\put(928,490){\circle{18}}
\put(952,490){\circle{18}}
\put(977,490){\circle{18}}
\put(1001,490){\circle{18}}
\put(1026,490){\circle{18}}
\put(1050,549){\circle{18}}
\put(1074,638){\circle{18}}
\put(1099,726){\circle{18}}
\put(1123,815){\circle{18}}
\put(1147,844){\circle{18}}
\put(1172,815){\circle{18}}
\put(1196,785){\circle{18}}
\put(1221,756){\circle{18}}
\put(1245,726){\circle{18}}
\put(1269,697){\circle{18}}
\put(1294,667){\circle{18}}
\put(1318,638){\circle{18}}
\put(1342,608){\circle{18}}
\put(1367,579){\circle{18}}
\put(1391,549){\circle{18}}
\put(1416,520){\circle{18}}
\end{picture}

%% file: fig4.tex
\setlength{\unitlength}{0.240900pt}
\ifx\plotpoint\undefined\newsavebox{\plotpoint}\fi
\sbox{\plotpoint}{\rule[-0.200pt]{0.400pt}{0.400pt}}%
\begin{picture}(1500,900)(0,0)
\font\gnuplot=cmr10 at 10pt
\gnuplot
\sbox{\plotpoint}{\rule[-0.200pt]{0.400pt}{0.400pt}}%
\put(201.0,122.0){\rule[-0.200pt]{4.818pt}{0.400pt}}
\put(181,122){\makebox(0,0)[r]{-0.5}}
\put(1420.0,122.0){\rule[-0.200pt]{4.818pt}{0.400pt}}
\put(201.0,196.0){\rule[-0.200pt]{4.818pt}{0.400pt}}
\put(181,196){\makebox(0,0)[r]{-0.4}}
\put(1420.0,196.0){\rule[-0.200pt]{4.818pt}{0.400pt}}
\put(201.0,269.0){\rule[-0.200pt]{4.818pt}{0.400pt}}
\put(181,269){\makebox(0,0)[r]{-0.3}}
\put(1420.0,269.0){\rule[-0.200pt]{4.818pt}{0.400pt}}
\put(201.0,343.0){\rule[-0.200pt]{4.818pt}{0.400pt}}
\put(181,343){\makebox(0,0)[r]{-0.2}}
\put(1420.0,343.0){\rule[-0.200pt]{4.818pt}{0.400pt}}
\put(201.0,417.0){\rule[-0.200pt]{4.818pt}{0.400pt}}
\put(181,417){\makebox(0,0)[r]{-0.1}}
\put(1420.0,417.0){\rule[-0.200pt]{4.818pt}{0.400pt}}
\put(201.0,491.0){\rule[-0.200pt]{4.818pt}{0.400pt}}
\put(181,491){\makebox(0,0)[r]{0}}
\put(1420.0,491.0){\rule[-0.200pt]{4.818pt}{0.400pt}}
\put(201.0,564.0){\rule[-0.200pt]{4.818pt}{0.400pt}}
\put(181,564){\makebox(0,0)[r]{0.1}}
\put(1420.0,564.0){\rule[-0.200pt]{4.818pt}{0.400pt}}
\put(201.0,638.0){\rule[-0.200pt]{4.818pt}{0.400pt}}
\put(181,638){\makebox(0,0)[r]{0.2}}
\put(1420.0,638.0){\rule[-0.200pt]{4.818pt}{0.400pt}}
\put(201.0,712.0){\rule[-0.200pt]{4.818pt}{0.400pt}}
\put(181,712){\makebox(0,0)[r]{0.3}}
\put(1420.0,712.0){\rule[-0.200pt]{4.818pt}{0.400pt}}
\put(201.0,785.0){\rule[-0.200pt]{4.818pt}{0.400pt}}
\put(181,785){\makebox(0,0)[r]{0.4}}
\put(1420.0,785.0){\rule[-0.200pt]{4.818pt}{0.400pt}}
\put(201.0,859.0){\rule[-0.200pt]{4.818pt}{0.400pt}}
\put(181,859){\makebox(0,0)[r]{0.5}}
\put(1420.0,859.0){\rule[-0.200pt]{4.818pt}{0.400pt}}
\put(201.0,122.0){\rule[-0.200pt]{0.400pt}{4.818pt}}
\put(201,81){\makebox(0,0){0}}
\put(201.0,839.0){\rule[-0.200pt]{0.400pt}{4.818pt}}
\put(325.0,122.0){\rule[-0.200pt]{0.400pt}{4.818pt}}
\put(325,81){\makebox(0,0){0.1}}
\put(325.0,839.0){\rule[-0.200pt]{0.400pt}{4.818pt}}
\put(449.0,122.0){\rule[-0.200pt]{0.400pt}{4.818pt}}
\put(449,81){\makebox(0,0){0.2}}
\put(449.0,839.0){\rule[-0.200pt]{0.400pt}{4.818pt}}
\put(573.0,122.0){\rule[-0.200pt]{0.400pt}{4.818pt}}
\put(573,81){\makebox(0,0){0.3}}
\put(573.0,839.0){\rule[-0.200pt]{0.400pt}{4.818pt}}
\put(697.0,122.0){\rule[-0.200pt]{0.400pt}{4.818pt}}
\put(697,81){\makebox(0,0){0.4}}
\put(697.0,839.0){\rule[-0.200pt]{0.400pt}{4.818pt}}
\put(820.0,122.0){\rule[-0.200pt]{0.400pt}{4.818pt}}
\put(820,81){\makebox(0,0){0.5}}
\put(820.0,839.0){\rule[-0.200pt]{0.400pt}{4.818pt}}
\put(944.0,122.0){\rule[-0.200pt]{0.400pt}{4.818pt}}
\put(944,81){\makebox(0,0){0.6}}
\put(944.0,839.0){\rule[-0.200pt]{0.400pt}{4.818pt}}
\put(1068.0,122.0){\rule[-0.200pt]{0.400pt}{4.818pt}}
\put(1068,81){\makebox(0,0){0.7}}
\put(1068.0,839.0){\rule[-0.200pt]{0.400pt}{4.818pt}}
\put(1192.0,122.0){\rule[-0.200pt]{0.400pt}{4.818pt}}
\put(1192,81){\makebox(0,0){0.8}}
\put(1192.0,839.0){\rule[-0.200pt]{0.400pt}{4.818pt}}
\put(1316.0,122.0){\rule[-0.200pt]{0.400pt}{4.818pt}}
\put(1316,81){\makebox(0,0){0.9}}
\put(1316.0,839.0){\rule[-0.200pt]{0.400pt}{4.818pt}}
\put(1440.0,122.0){\rule[-0.200pt]{0.400pt}{4.818pt}}
\put(1440,81){\makebox(0,0){1}}
\put(1440.0,839.0){\rule[-0.200pt]{0.400pt}{4.818pt}}
\put(201.0,122.0){\rule[-0.200pt]{298.475pt}{0.400pt}}
\put(1440.0,122.0){\rule[-0.200pt]{0.400pt}{177.543pt}}
\put(201.0,859.0){\rule[-0.200pt]{298.475pt}{0.400pt}}
\put(41,490){\makebox(0,0){${\rho \langle v \rangle}$}}
\put(820,40){\makebox(0,0){${\rho}$}}
\put(201.0,122.0){\rule[-0.200pt]{0.400pt}{177.543pt}}
\put(226,461){\circle{18}}
\put(251,432){\circle{18}}
\put(275,402){\circle{18}}
\put(300,373){\circle{18}}
\put(325,343){\circle{18}}
\put(350,314){\circle{18}}
\put(374,284){\circle{18}}
\put(399,255){\circle{18}}
\put(424,225){\circle{18}}
\put(449,196){\circle{18}}
\put(474,166){\circle{18}}
\put(498,137){\circle{18}}
\put(523,135){\circle{18}}
\put(548,165){\circle{18}}
\put(573,194){\circle{18}}
\put(597,225){\circle{18}}
\put(622,255){\circle{18}}
\put(647,284){\circle{18}}
\put(672,314){\circle{18}}
\put(697,343){\circle{18}}
\put(721,373){\circle{18}}
\put(746,402){\circle{18}}
\put(771,432){\circle{18}}
\put(796,461){\circle{18}}
\put(821,490){\circle{18}}
\put(845,520){\circle{18}}
\put(870,549){\circle{18}}
\put(895,579){\circle{18}}
\put(920,608){\circle{18}}
\put(944,638){\circle{18}}
\put(969,667){\circle{18}}
\put(994,697){\circle{18}}
\put(1019,726){\circle{18}}
\put(1044,756){\circle{18}}
\put(1068,785){\circle{18}}
\put(1093,815){\circle{18}}
\put(1118,844){\circle{18}}
\put(1143,844){\circle{18}}
\put(1167,815){\circle{18}}
\put(1192,785){\circle{18}}
\put(1217,756){\circle{18}}
\put(1242,726){\circle{18}}
\put(1267,697){\circle{18}}
\put(1291,667){\circle{18}}
\put(1316,638){\circle{18}}
\put(1341,608){\circle{18}}
\put(1366,579){\circle{18}}
\put(1390,549){\circle{18}}
\put(1415,520){\circle{18}}
\end{picture}

%% file: fig5.tex
\setlength{\unitlength}{0.240900pt}
\ifx\plotpoint\undefined\newsavebox{\plotpoint}\fi
\sbox{\plotpoint}{\rule[-0.200pt]{0.400pt}{0.400pt}}%
\begin{picture}(1500,900)(0,0)
\font\gnuplot=cmr10 at 10pt
\gnuplot
\sbox{\plotpoint}{\rule[-0.200pt]{0.400pt}{0.400pt}}%
\put(221.0,122.0){\rule[-0.200pt]{4.818pt}{0.400pt}}
\put(201,122){\makebox(0,0)[r]{-0.15}}
\put(1420.0,122.0){\rule[-0.200pt]{4.818pt}{0.400pt}}
\put(221.0,245.0){\rule[-0.200pt]{4.818pt}{0.400pt}}
\put(201,245){\makebox(0,0)[r]{-0.1}}
\put(1420.0,245.0){\rule[-0.200pt]{4.818pt}{0.400pt}}
\put(221.0,368.0){\rule[-0.200pt]{4.818pt}{0.400pt}}
\put(201,368){\makebox(0,0)[r]{-0.05}}
\put(1420.0,368.0){\rule[-0.200pt]{4.818pt}{0.400pt}}
\put(221.0,491.0){\rule[-0.200pt]{4.818pt}{0.400pt}}
\put(201,491){\makebox(0,0)[r]{0}}
\put(1420.0,491.0){\rule[-0.200pt]{4.818pt}{0.400pt}}
\put(221.0,613.0){\rule[-0.200pt]{4.818pt}{0.400pt}}
\put(201,613){\makebox(0,0)[r]{0.05}}
\put(1420.0,613.0){\rule[-0.200pt]{4.818pt}{0.400pt}}
\put(221.0,736.0){\rule[-0.200pt]{4.818pt}{0.400pt}}
\put(201,736){\makebox(0,0)[r]{0.1}}
\put(1420.0,736.0){\rule[-0.200pt]{4.818pt}{0.400pt}}
\put(221.0,859.0){\rule[-0.200pt]{4.818pt}{0.400pt}}
\put(201,859){\makebox(0,0)[r]{0.15}}
\put(1420.0,859.0){\rule[-0.200pt]{4.818pt}{0.400pt}}
\put(221.0,122.0){\rule[-0.200pt]{0.400pt}{4.818pt}}
\put(221,81){\makebox(0,0){0}}
\put(221.0,839.0){\rule[-0.200pt]{0.400pt}{4.818pt}}
\put(343.0,122.0){\rule[-0.200pt]{0.400pt}{4.818pt}}
\put(343,81){\makebox(0,0){0.1}}
\put(343.0,839.0){\rule[-0.200pt]{0.400pt}{4.818pt}}
\put(465.0,122.0){\rule[-0.200pt]{0.400pt}{4.818pt}}
\put(465,81){\makebox(0,0){0.2}}
\put(465.0,839.0){\rule[-0.200pt]{0.400pt}{4.818pt}}
\put(587.0,122.0){\rule[-0.200pt]{0.400pt}{4.818pt}}
\put(587,81){\makebox(0,0){0.3}}
\put(587.0,839.0){\rule[-0.200pt]{0.400pt}{4.818pt}}
\put(709.0,122.0){\rule[-0.200pt]{0.400pt}{4.818pt}}
\put(709,81){\makebox(0,0){0.4}}
\put(709.0,839.0){\rule[-0.200pt]{0.400pt}{4.818pt}}
\put(830.0,122.0){\rule[-0.200pt]{0.400pt}{4.818pt}}
\put(830,81){\makebox(0,0){0.5}}
\put(830.0,839.0){\rule[-0.200pt]{0.400pt}{4.818pt}}
\put(952.0,122.0){\rule[-0.200pt]{0.400pt}{4.818pt}}
\put(952,81){\makebox(0,0){0.6}}
\put(952.0,839.0){\rule[-0.200pt]{0.400pt}{4.818pt}}
\put(1074.0,122.0){\rule[-0.200pt]{0.400pt}{4.818pt}}
\put(1074,81){\makebox(0,0){0.7}}
\put(1074.0,839.0){\rule[-0.200pt]{0.400pt}{4.818pt}}
\put(1196.0,122.0){\rule[-0.200pt]{0.400pt}{4.818pt}}
\put(1196,81){\makebox(0,0){0.8}}
\put(1196.0,839.0){\rule[-0.200pt]{0.400pt}{4.818pt}}
\put(1318.0,122.0){\rule[-0.200pt]{0.400pt}{4.818pt}}
\put(1318,81){\makebox(0,0){0.9}}
\put(1318.0,839.0){\rule[-0.200pt]{0.400pt}{4.818pt}}
\put(1440.0,122.0){\rule[-0.200pt]{0.400pt}{4.818pt}}
\put(1440,81){\makebox(0,0){1}}
\put(1440.0,839.0){\rule[-0.200pt]{0.400pt}{4.818pt}}
\put(221.0,122.0){\rule[-0.200pt]{293.657pt}{0.400pt}}
\put(1440.0,122.0){\rule[-0.200pt]{0.400pt}{177.543pt}}
\put(221.0,859.0){\rule[-0.200pt]{293.657pt}{0.400pt}}
\put(41,490){\makebox(0,0){${\rho \langle v \rangle}$}}
\put(830,40){\makebox(0,0){${\rho}$}}
\put(221.0,122.0){\rule[-0.200pt]{0.400pt}{177.543pt}}
\put(245,443){\circle{18}}
\put(270,400){\circle{18}}
\put(294,363){\circle{18}}
\put(319,327){\circle{18}}
\put(343,292){\circle{18}}
\put(367,269){\circle{18}}
\put(392,244){\circle{18}}
\put(416,225){\circle{18}}
\put(440,211){\circle{18}}
\put(465,191){\circle{18}}
\put(489,194){\circle{18}}
\put(514,194){\circle{18}}
\put(538,188){\circle{18}}
\put(562,207){\circle{18}}
\put(587,209){\circle{18}}
\put(611,238){\circle{18}}
\put(635,267){\circle{18}}
\put(660,284){\circle{18}}
\put(684,310){\circle{18}}
\put(709,322){\circle{18}}
\put(733,352){\circle{18}}
\put(757,396){\circle{18}}
\put(782,426){\circle{18}}
\put(806,471){\circle{18}}
\put(831,499){\circle{18}}
\put(855,517){\circle{18}}
\put(879,548){\circle{18}}
\put(904,602){\circle{18}}
\put(928,632){\circle{18}}
\put(952,656){\circle{18}}
\put(977,682){\circle{18}}
\put(1001,692){\circle{18}}
\put(1026,735){\circle{18}}
\put(1050,746){\circle{18}}
\put(1074,757){\circle{18}}
\put(1099,764){\circle{18}}
\put(1123,762){\circle{18}}
\put(1147,759){\circle{18}}
\put(1172,788){\circle{18}}
\put(1196,779){\circle{18}}
\put(1221,755){\circle{18}}
\put(1245,758){\circle{18}}
\put(1269,728){\circle{18}}
\put(1294,712){\circle{18}}
\put(1318,695){\circle{18}}
\put(1342,654){\circle{18}}
\put(1367,626){\circle{18}}
\put(1391,583){\circle{18}}
\put(1416,538){\circle{18}}
\end{picture}